
\input phyzzx
\unnumberedchapters
\pubnum{$\caps ROM2F-92/53$}
\titlepage
\title{New \ \ Developments \ \ in \ \
Open - String \ \ Theories}
\author{Gianfranco \ \ Pradisi \ \ \ and \ \ \ Augusto \ \
Sagnotti}
\address{Dipartimento di Fisica \break
Universit\`a di Roma ``Tor Vergata'' \break
I.N.F.N., Sezione di Roma ``Tor Vergata'' \break
Via della Ricerca Scientifica, 1 \break
00133 Roma  ITALY}
\vskip .6 truein
\abstract
{\baselineskip=14pt
The study of string models including both unoriented closed
strings and open strings presents a number of new features when compared
to the standard case of models of oriented closed strings only. We review
some basic features of the construction of these models, describing in
particular how gauge symmetry breaking can be achieved in this case. We
also review some peculiar properties of the Green-Schwarz anomaly
cancellation mechanism that present themselves in lower-dimensional
open-string models.}
\vskip 1.5 truein
\centerline{\sl Talk presented at the Tenth National General Relativity
Conference}
\centerline{\sl Bardonecchia, ITALY, September 1992}
\endpage
\pagenumber=2
\vskip 24pt

The study of models of {\it oriented} closed strings has been the object of
most of the research carried out in String Theory over the last few
years \Ref\clos{
A.M. Polyakov, {\sl Phys. Lett.} {\bf 103B} (1981) 207, 211; \nextline
D. Friedan, {\it in} Les Houches '82, ``Recent Advances in
Field Theory and \nextline Statistical Mechanics'', eds. J.B. Zuber and R.
Stora (Elsevier, 1984);\nextline
A.A. Belavin, A.M. Polyakov and A.B. Zamolodchikov, {\sl Nucl. Phys.}
{\bf B241} (1984) 333;\nextline
D. Friedan, E. Martinec and S. Shenker, {\sl Nucl. Phys.} {\bf B271} (1986)
93;\nextline
D. Gross, J.A. Harvey, E. Martinec and R. Rohm, {\sl Nucl. Phys.}
{\bf B256} (1985) 253.}
\Ref\closed{L. Dixon, J.A. Harvey, C. Vafa and E. Witten,
{\sl Nucl. Phys.} {\bf B261} (1985) 678; \nextline {\bf B 274} (1986)
285;\nextline L. Dixon, D. Friedan, E. Martinec and S. Shenker, {\sl Nucl.
Phys.} {\bf B282} (1987) 13; \nextline S. Hamidi and C. Vafa, {\sl Nucl.
Phys.} {\bf B279} (1987) 465.} \Ref\closedd{I. Antoniadis, C. Bachas and C.
Kounnas, {\sl Nucl. Phys.} {\bf B288} (1987) 87; \nextline
H. Kawai, D.C. Lewellen and S.H. Tye, {\sl Nucl. Phys.} {\bf B288} (1987) 1;
\nextline
W. Lerche, D. L\"ust and A.N. Schellekens, {\sl Nucl. Phys.} {\bf B287}
(1987) 477; \nextline
K.S. Narain, M.H. Sarmadi and C. Vafa, {\sl Nucl. Phys.} {\bf B288}
(1987) 551.}.  Two key concepts have emerged. The first one is the {\it
(super)conformal invariance } [\clos]~of the world-sheet theory describing the
degrees of freedom of the string, a
remnant, after gauge fixing, of the local invariances of the models.  The
second is {\it modular invariance}\Ref\mod{J.
Shapiro, {\sl Phys. Rev.} {\bf D5} (1972) 1945; \nextline W. Nahm, {\sl
Nucl. Phys.} {\bf B114} (1976) 174; \nextline F. Gliozzi, D. Olive and J.
Scherk, {\sl Nucl. Phys.} {\bf B122} (1977) 253.}, a subtle global
constraint  deeply linked to the very nature of oriented closed strings.  The
meaning of this constraint may be simply appreciated by referring to the
torus displayed in fig. 1.
In sharp contrast with the case of particles, when oriented closed
strings propagate there is generally {\it no} unique notion of ``time''
direction on the world sheet. The ambiguity, however, is harmless, as long
as the models are endowed with a corresponding symmetry.  The symmetry is
precisely modular invariance, and for the torus one finds the infinite group
$SL(2,Z)$ generated by the two transformations
$$
\tau \ \rightarrow \ \tau \ + \ 1	\qquad {\rm and} \qquad  \tau \ \rightarrow
\  - \ {1 \over \tau}	\qquad .
\eqn\mod
$$
The torus amplitude is very important, since it contains a lot of
information about the spectrum of the models.  Thus, the modular invariance
of the torus amplitude is
a basic criterion to identify consistent models of oriented closed strings.
The role of modular transformations in higher-genus amplitudes is also quite
interesting.  In particular, the invariance at genus two may be linked to the
relation between spin and statistics [\closedd].

The extension of these results to models including both unoriented closed
strings and open strings (referred to, for brevity, as open-string
models) has proved  harder than was naively expected.  There are actually two
reasons for this.  First of all, the propagation of unoriented and/or open
strings reduces in part (or, at times, completely) the ambiguity in the
choice of time direction on the world sheet.  As a result, the role
of modular transformations in these models need be reconsidered.
Moreover, the end points of open strings are special, and they may be
associated to the internal (Chan-Paton) symmetry \Ref\cp{H.M. Chan and J.E.
Paton, {\sl Nucl. Phys.} {\bf B10} (1969) 516;\nextline
J.H. Schwarz, in ``Current Problems in Particle Physics'', \nextline
{\sl Proc. Johns Hopkins} {\bf 6} (Firenze, 1982);\nextline
N. Marcus and A. Sagnotti, {\sl Phys. Lett.} {\bf 119B} (1982) 97.} of the
models.

Referring to fig. 2, let us consider the propagation on an annulus of open
strings with end charges possibly of two different kinds.  This phenomenon
corresponds to choosing a ``vertical'' time direction on the world sheet.
On the other hand, if a ``horizontal'' time direction is chosen, for instance
by performing a modular transformation on the previous setting, the picture
changes dramatically: now one sees closed strings propagating along a tube
with two boundaries at its ends.  The lesson is therefore that, though quite
important, modular transformations are {\it not} an invariance of these
models. Rather, they link different sectors of the same model.  In particular,
open strings {\it need} closed strings for their consistency, since their
propagation {\it is}, in some respect, the propagation of closed strings.
This crucial fact was pointed out long ago by Lovelace \Ref\lov{
C. Lovelace, {\sl Phys. Lett.} {\bf 134B} (1970) 703.}.
What should we learn from these
considerations ?  First of all, the structure of models of oriented closed
strings only is simpler than that of open-string models.  Thus, reverting the
original observation of Lovelace, it would appear quite convenient to use
models of oriented closed strings as the starting point in the process of
model building for open-string theories.  The suitable models should allow,
in particular, for proper reflections at the ends of the tube that, as such,
mix left-moving and right-moving waves. Thus, the natural condition is to
require a symmetry in the spectrum under the interchange of these two kinds
of waves.  We shall call ``left-right symmetric'' the closed-string models in
this class  \Ref\car{A. Sagnotti, {\it in} Cargese '87, ``Non-Perturbative
Quantum Field Theory'' \nextline  (Plenum Press, New York, 1988).}.

 As to model building, it is natural to begin by associating to the simplest
known (left-right symmetric) models of oriented closed strings, the closed
bosonic string in $D=26$ and the type-IIb superstring in $D=10$, corresponding
(classes of) open-string models, that we shall call their ``descendants''.  The
$D=10$ case is well known, of course, the open-string model being the $SO(32)$
type-I superstring of Green and Schwarz\Ref\gs{M.B. Green and J.H. Schwarz,
{\sl Phys. Lett.} {\bf 149B} (1984) 117;\nextline {\bf 151B} (1985) 21.}.
The novelty of this viewpoint is the connection between this model and the
type-IIb superstring.  In a similar fashion, in $D=26$ one encounters a
peculiar bosonic open-string model, with gauge group $SO(8192)$, first
described in ref. \Ref\sobig{M.R. Douglas and B. Grinstein,
{\sl Phys. Lett.} {\bf 183B} (1987) 52;\nextline
S. Weinberg, {\sl Phys. Lett.} {\bf 187B} (1987) 287;\nextline
N. Marcus and A. Sagnotti, {\sl  Phys. Lett.} {\bf 188B} (1987) 58.}.
Clearly, the issue is how to associate to a generic (left-right symmetric)
closed-string model a (class of) open-string models, and in this respect it
is quite illuminating to take a critical look at the relation in these
simple cases.

Referring to figure 3, let us note that, whereas in the bosonic closed string
the genus-one contribution to the vacuum energy is fully described by the
torus amplitude, in the $SO(8192)$ model there are three
additional contributions.  These may be associated to the other three surfaces
of vanishing Euler number that contain holes and/or cross-caps, the
Klein bottle, the annulus and the M\"obius strip.  These surfaces are quite
different: the first one has no boundaries and is not orientable, the second
one has two boundaries and is orientable, and the third one has one boundary
as is, again, not orientable. Their role is also quite distinct: the first
surface completes the symmetrization of the closed-string spectrum under the
interchange of left and right modes, and its contribution is
$$
K \ = \ {1 \over 2} \int_0^{\infty}  {{d \tau} \over {{\tau}^{14}}} \ {1 \over
{\big({\eta (2 i \tau )}\big)^{24}}}	\qquad .
\eqn\klbig
$$
On the other hand, the annulus and the M\"obius
strip describe the open-string sector, properly symmetrized under the
interchange of the charges at the ends of the open string.  Their
contributions are
$$
A \ = \ {N^2 \over 2} \int_0^{\infty}  {{d \tau} \over {{\tau}^{14}}} \ {1
\over {\big({\eta ( i \tau
/ 2 )}\big)^{24}}}
\eqn\klbig
$$
and
$$
M \ = \ - \ {N \over 2} \int_0^{\infty}  {{d \tau} \over {{\tau}^{14}}} \ {1
\over {\big({\eta ( {i \tau}/2 + 1/2)}\big)^{24}}}	\qquad .
\eqn\klbig
$$

All this is somewhat reminiscent of the familiar structure
that one encounters when passing from a closed-string model to a more
complicated one via an orbifold construction [\closed],
whatever the geometrical interpretation of the latter.  Namely, one starts
by restricting the original spectrum by a suitable projection, and then adds
new (twisted) sectors with rather peculiar features.  In the closed-string
case, when starting from a (target-space) torus,
the additional strings are closed only on the (target-space) orbifold but not
on its covering torus. On the other hand, in this open-string example, the
open strings are closed only on the double covers of the relevant Riemann
surfaces.  In this respect, open-string models should be regarded as
``parameter-space orbifolds'' of their ``parent'' closed-string models
[\car].  In models of oriented closed strings, modular invariance demands
that one adds the twisted sector to the spectrum. The issue is then what
fixes the (twisted) open-string sector of open-string models.

The answer to this crucial question may be described quite neatly by
referring to fig.~4.  There the three additional surfaces of fig. 3 are
described in terms of a choice of time direction that corresponds to the
propagation of closed strings.  With due care, this choice has the
virtue of making the three contributions of Klein bottle, annulus and
M\"obius strip comparable.  Still, there is a crucial
difference between these and the torus amplitude.
Whereas the modular invariance of the
torus amplitude allows one to restrict the integration region, while leaving
out the  ``ultraviolet'' line $Im( \tau ) = 0$, the lack of modular
invariance in the last three surfaces makes their Teichm\"uller
spaces identical to their moduli spaces. All coincide with
(translates of) the imaginary axis, and thus all include the ``ultraviolet''
point $Im( \tau ) = 0$.  The ultraviolet singularity may now be dealt with by
considering suitable combinations of the contributions from the three
surfaces, to be written in the vacuum channel. In this
respect, one is actually cancelling an infrared divergence that arises in the
limit of very long tubes. The idea, following ref. [\gs], is
to express all amplitudes in terms of $q = exp( - 2 \pi \tau )$,
where $\tau$ denotes the ``time'' measured on the double covers along the
tube, and to arrive at a
``principal-part'' prescription for the resulting pole part at $q=0$.
In the bosonic string one finds
$$
{\tilde{Z}}_{tot}\ = \ - \ {N \over 2} \ \int_0^1 \
{dq \over {2 \pi q}} \ {1 \over {\eta ( - q )}^{24}} \
+ \ {{N^2 + 2^{26}} \over {4 \ 2^{13}}} \ \int_0^1 \
{dq \over {2 \pi q}} \ {1 \over {\eta ( q )}^{24}} \qquad ,
\eqn\ztilbos
$$
and for $N = 2^{13}$, corresponding to the $SO(8192)$ model [\sobig],
$$
{\tilde{Z}}_{tot}\ = {{2^{13}} \over 2} \ PP \ \int_{-1}^1 \
{dq \over {2 \pi q}} \ {1 \over {\eta ( q )}^{24}}	\qquad  .
\eqn\zprinc
$$
This choice disposes of the massless tadpole, while the additional singularity
introduced by the tachyon is not regulated.

In general, the ``tadpole
conditions'' fix (part of) the dimensionalities of the allowed
open-string gauge groups.  These appear in a multiplicative fashion in the
annulus and M\"obius amplitudes, being linked to the degeneracy present in
the definition of the boundaries.  Upon factorization, the resulting
conditions may then be related to suitable expectation values on the two
``genus-one-half'' surfaces, the disk and the projective plane\Ref\bsbig{M.
Bianchi and A. Sagnotti, {\sl Phys. Lett.} {\bf 211B} (1988) 411.}(fig. 5).
For instance, in the bosonic model one finds for the total pole part
$$
Z_{s.p.} \ = \ 24 {{(N - 2^{13})}^2 \over {2^{13}}} \
\int_0^1 \ {dq \over {2 \pi q}}	\qquad ,
\eqn\zsp
$$
and from this one may deduce, by factorization, the total ``genus-one-half''
tadpole [\bsbig]
$$
\Gamma \ = \ 2^8 \ \sqrt{3 \pi} \ \bigg| {N \over {2^{13}}} \ - \ 1 \bigg|
\qquad .
 \eqn\tad
$$
More refined considerations \Ref\pc{J. Polchinski and Y. Cai, {\sl
Nucl. Phys.} {\bf B296} (1988) 91.} show that, in models with chiral fermions,
some sectors contribute ${\it unphysical}$ modes that may flow along
the tubes.  The tadpoles of these unphysical modes signal the
presence of gauge and gravitational anomalies.

Let us illustrate the key features of the construction in the simplest of
all possible new settings, by working out the open-string descendants of
the left-right symmetric models in ten dimensions.  To this end, we begin by
recalling the definition of the theta constant with characteristics $\alpha$
and $\beta$,
$$
\theta {\alpha \brack \beta} \ = \
\sum_{n=0}^{\infty} \ e^{i \pi \tau ( n + \alpha / 2 )^2 \ + \ 2 \pi i (
n + \alpha / 2 ) \beta / 2} \qquad .
 \eqn\thetas
$$
Starting from this expression, one may define the level-one $SO(2n)$
characters,
$$
\eqalign{ O_{2n} \ &= \ {1 \over {2 \ {\eta}^n}} \ \bigg( {\theta}^n {0
\brack 0} \ + \ {\theta}^n {0 \brack 1/2} \bigg) \qquad , \cr
V_{2n} \ &= \ {1 \over {2 \ {\eta}^n}} \ \bigg( {\theta}^n {0
\brack 0}
\ - \ {\theta}^n {0 \brack 1/2} \bigg)  \qquad , \cr
S_{2n} \ &= \ {1 \over {2 \ {\eta}^n}} \ \bigg( {\theta}^n {1/2 \brack 0} \ +
\ i^n \ {\theta}^n {1/2 \brack 1/2} \bigg) \qquad , \cr
C_{2n} \ &= \ {1
\over {2 \ {\eta}^n}} \ \bigg( {\theta}^n {1/2 \brack 0} \ - \
i^n \  {\theta}^n {1/2 \brack 1/2} \bigg) \qquad , \cr}
\eqn\levone
$$
where $\eta$ is the Dedekind function. These four characters correspond to the
four conjugacy classes of $SO(8)$ representations, and owe their names to the
lowest-dimensional representations in these classes.  These, in their turn,
characterize the lowest-mass states in the corresponding sectors.  Thus,
$O_8$ starts with a scalar (a tachyon), while $V_8$, $S_8$ and $C_8$
start with massless states, a vector and two conjugate spinors, respectively.
On this basis of characters, the two generating transformations of eq. \mod ~
are represented by the two matrices
$$
S \ = \ {1 \over 2} \
\pmatrix{1&1&1&1\cr1&1&-1&-1\cr1&-1&1&-1\cr1&-1&-1&1\cr}		\qquad {\rm and}
\qquad
T \ = \ exp( - \ {{ 2 i \pi} \over 3} ) \
\pmatrix{1&0&0&0\cr0&-1&0&0\cr0&0&-1&0\cr0&0&0&-1\cr} \quad .
\eqn\matrices
$$
In this notation, and leaving aside the contributions from the transverse
bosons and from the measure over the moduli, the partition functions of the
ten-dimensional type-II models of ref. \Ref\dh{ L. Dixon and J.A. Harvey,
{\sl Nucl. Phys.} {\bf B274} (1986) 93;\nextline N. Seiberg and E. Witten,
{\sl Nucl. Phys.} {\bf B276} (1986) 272.} are
$$
\eqalign{ T_{IIA} \ &= \ ( V_8 \ - \ S_8 ) ({ \bar{V}}_8 \ - \ { \bar{C}}_8
) \qquad , \cr
T_{IIB} \ &= \ {| V_8 \ - \ S_8 |}^2 \qquad , \cr
T_{0A} \ &= \ {| O_8 |}^2 \ + \ {| V_8 |}^2 \ + \ S_8 \ {\bar{C}}_8 \ + \
C_8 \ {\bar{S}}_8 \qquad , \cr
T_{0B} \ &= \ {| O_8 |}^2 \ + \ {| V_8 |}^2 \ + \ {| S_8 |}^2 \ + \
{| C_8 |}^2 \qquad . \cr}
\eqn\chars
$$
The descendant of the type-IIb theory is the type-I $SO(32)$ model of Green
and Schwarz.  On the other hand, the type-IIa model is not left-right
symmetric, and thus is not suitable for our construction.  We are thus left
with the remaining two non-supersymmetric models.  Though tachyonic, they
are an amusing testing ground for these ideas, since their open-string
descendants contain a number of different sectors.
In the following, we construct their descendants simultaneously, in order to
elicit the differences between the two cases.

The first step is writing the Klein bottle amplitude in the direct channel.
This contribution is meant to complete the symmetrization of the $NS-NS$
sector of the closed string and the antisymmetrization of the $R-R$
sector of the closed string under the interchange of left and right
modes\foot{As for the ``spin-statistics'' signs of the first of
refs. [\closedd], the need to
antisymmetrize the $R-R$ sector may be traced to the behavior of the
gravitino determinant at genus two\Ref\mod{ M. Bianchi and A. Sagnotti, {\sl
Phys. Lett.} {\bf B231} (1989) 389.}.}.  One finds
$$
\eqalign{K_{0A} \ &= \ {1 \over 2} \ ( O_8 \ + \ V_8 ) \qquad , \cr
K_{0B} \ &= \ {1 \over 2} \ ( O_8 \ + \ V_8 \ - \ S_8 \ - \ C_8 ) \qquad ,
\cr}
\eqn\klein
$$
where the argument is $2 i \tau$, with $\tau$ the ``proper time'' of the
closed string.
In the vacuum channel these expressions become
$$
\eqalign{{\tilde{K}}_{0A} \ &= \ {{2^5} \over 2} \ ( O_8 \ + \ V_8 )
\qquad , \cr {\tilde{K}}_{0B} \ &= \ {{2^6} \over 2} \ V_8
\qquad . \cr}
\eqn\kleint
$$
The powers of two are introduced by the measure over the moduli, once
these amplitudes are expressed in terms of the Teichm\"uller parameters of
their double covers, the natural choice if one is to compare
contributions from all three additional surfaces.

The second, more difficult, step, is the contruction\Ref\bs{M. Bianchi and A.
Sagnotti, {\sl Phys. Lett.} {\bf 247B} (1990)
517.}\Ref\bss{M. Bianchi and A. Sagnotti, {\sl Nucl. Phys.} {\bf B361} (1991)
519.} of the annulus amplitude.  The starting point is, again, the
closed-string spectrum, or rather the portion allowed in the tube by the holes
at its ends. The rule for the allowed sectors is quite simple in this case :
they are precisely the ones that fuse into the identity of the fusion
algebra\foot{In this case $V_8$, due to the Minkowski signature of the
ten-dimensional space time! [\closedd]} with their anti-holomorphic partners
in the closed-string GSO projection.  This is precisely the condition that
boundaries respect the transverse Lorentz group, $SO(8)$, a {\it
local} symmetry of these models.  In lower dimensions, one may relax this
condition to define models where boundaries respect
only part of the symmetries of the ``parent'' closed string.  For
instance, one may extend the familiar toroidal construction of ref. \Ref\nsw{
K.S. Narain, {\sl Phys. Lett.} {\bf 169B} (1986) 41;\nextline K.S. Narain,
M.H. Sarmadi and E. Witten, {\sl Nucl. Phys.} {\bf B279} (1987) 369.}.
The additional freedom is then related to marginal deformations of
the conformal theory that draw their origin directly from the open-string
sector \Ref\torus{M. Bianchi, G. Pradisi and A. Sagnotti, {\sl  Nucl. Phys.}
{\bf B376} (1992) 365.}. Returning to our $D=10$ models, it is simple to see
that, since all $SO(8)$ representations are self-conjugate, in the $0A$ models
only $O_8$ and $V_8$ can flow in the vacuum channel, while in the $0B$ models
all four characters can flow (figure 6).

The next, crucial observation, is that one still has the freedom
to choose the reflection coefficients for the various sectors.  Thus, one
may write
$$
{\tilde{A}}_{0A} \ = \ {{2^{-5}} \over 2} \ \big(
{( n_B  +  n_F )}^2 \ V_8 \ + \ {( n_B  -  n_F )}^2 \ O_8 \big)
\eqn\atildeoa
$$
and
$$
\eqalign{
{\tilde{A}}_{0B} \ &= \ {{2^{-6}} \over 2} \ \big(
{( n_o  +  n_v  +  n_s  +  n_c )}^2 \ V_8 \ + \
{( n_o  +  n_v  -  n_s  -  n_c )}^2 \ O_8 \ + \ \cr
&( n_o  -  n_v  +  n_s  -  n_c )^2 \ S_8 \ + \
{( n_o  -  n_v  -  n_s  +  n_c )}^2 \ C_8 \big) \qquad , \cr}
\eqn\atildeob
$$
where the argument is $i \tau$, the Teichm\"uller parameter of the
double covers. It should be appreciated that the coefficients in these
amplitudes are {\it perfect squares}.  This reflects the fact that the
individual sectors of the spectrum flow independently along the tube, while
undergoing two reflections at its ends.
In the direct (open-string) channel, these amplitudes
become
$$
A_{0A} \ = \ {{{n_B}^2  +  {n_F}^2} \over 2} \big( O_8 \ +
 \ V_8 \big) \
- \  n_B  \ n_F \big( S_8 \ + \ C_8 \big)
\eqn\aoa
$$
and
$$
\eqalign{ A_{0B} \ = \ {{{n_o}^2 +  {n_v}^2 + {n_s}^2 +  {n_c}^2 }
\over 2} \ &V_8 \ + \ ( n_o n_v +  n_s n_c ) O_8 \ - \cr
&( n_o n_c \ + \ n_v n_s ) S_8 \ - \
( n_o n_s \ + \ n_v n_c ) C_8
\qquad , \cr}
\eqn\aob
$$
where the argument is $ {i \tau} \over 2 $, with $\tau$ the ``proper
time'' of open strings.

The final step is the construction of the M\"obius amplitude.  The guidance
comes, again, from the transverse channel.  The basic observation is
that the M\"obius strip may be represented as a tube with
a hole and a cross-cap at its ends, and as
such it may accommodate {\it all the characters common to both $\tilde{A}$
and $\tilde{K}$}.  Moreover, apart from signs, the corresponding reflection
coefficients are geometric means of those in  $\tilde{A}$ and $\tilde{K}$.
Thus\foot{As in refs. [\bs] and [\bss], we are using a real basis of M\"obius
characters.}
$$
 {\tilde{M}}_{0A} \ = \ - \ \big(
{( n_B  +  n_F )} \ \hat{V}_8 \ + \
{( n_B \ - \ n_F )} \ \hat{O}_8
\big) \eqn\mtildeoa
$$
and
$$
{\tilde{M}}_{0B} \ = \ {( n_o  +  n_v  +  n_s
+ n_c )} \ \hat{V}_8 \qquad ,
\eqn\matildeob
$$
where the argument is  $\big( {i \tau} + 1 / 2 \big)$, the Teichm\"uller
parameter of the double covers. The
modular transformation to the direct channel is then effected by the matrix
$$
P \ = \ {T^{1 \over 2}} \ S \ T^2 \ S \ {T^{1 \over 2}} \qquad ,
\eqn\matp
$$
and the resulting projections of the open-string spectrum are
$$
M_{0A} \ = \ - \ {1 \over 2} \ \big( ( n_B \ + \ n_F ) \
\hat{V}_8  \ + \  ( n_B \ - \ n_F ) \
\hat{O}_8 \big) \eqn\maoa
$$
and
$$
M_{0B} \ =  - \ {1 \over 2} \ ( n_o \ + \ n_v \ + \
n_s \ + \ n_c ) \ \hat{V}_8 \qquad ,
\eqn\mob
$$
where the argument is $\big( {i \tau} / 2  + 1 / 2 \big)$, with $\tau$
the``proper time'' of open strings.
These expressions are indeed the proper symmetrizations of those
in eqs. \aoa~ and \aob.

The last step is then imposing the tadpole conditions.  In the $0A$ models
there is only one tadpole condition, coming from the $V_8$ sector, and
the result is
$$
n_B \ + \ n_F \ = \ 32	\qquad .
\eqn\tadoa
$$
On the other hand, in the $0B$ models there are three tadpole conditions,
coming from the sectors $V_8$, $S_8$ and $C_8$, that lead to
$$
n_o \ + \ n_v \ + \ n_s \ + \ n_c \ = 64 \qquad ,
\eqn\tadb
$$
$$
n_o \ = \ n_v \qquad {\rm and} \qquad n_s \ = \ n_c \qquad .
\eqn\tadbb
$$

The presence of the additional conditions in the $OB$ case is particularly
gratifying, since these open-string models are chiral, and eqs. \tadbb~are
sufficient to ensure the cancellation of all anomalies.  Strictly speaking, by
imposing tadpole conditions one is eliminating the irreducible part of the
anomaly polynomial.  The antisymmetric
tensors in the models then  dispose of the non-irreducible traces, via a
(generalized) Green-Schwarz mechanism \Ref\as{A. Sagnotti, {\sl Phys. Lett.}
{\bf 294B} (1992) 196.}.  In order to appreciate these remarks, let us first
pause briefly to describe the meaning of these expressions.

The first observation is that the annulus amplitude is a
second-degree polynomial in the Chan-Paton multiplicities.  This is
precisely as it should be, since open strings carry a pair of charges
at their ends.  Referring for simplicity to the 0A model, we see that if
$n_B =32$ there is a single open-string sector, with matter content
corresponding to $( O_8 + V_8 )$, and {\it all} open strings carry a pair of
identical charges.  Thus, they {\it all} flow in the M\"obius strip, and the
gauge group is $SO(32)$. In agreement with ref. [\cp], one is filling up
complete matrices, apart from the (anti)symmetrizations allowed by the
``twist'' symmetry.  If, on the other hand, $n_B$ is not equal to $32$, there
are additional open strings, with matter content
corresponding to $( S_8 + C_8 )$ and, at their ends, two charges of
different types, with multiplicities $n_B$ and $n_F$ respectively. All
this has
a nice interpretation, described in fig. 7 : upon symmetry breaking, the
``missing'' Chan-Paton polarizations are simply {\it moved} to new
sectors of the spectrum.  If this is done compatibly with the fusion rules,
the
factorization constraints are still satisfied.  This brings us to another
interesting issue, that has to do with the fusion rules themselves.
Namely,
the very form of eqs. \aoa ~ and \aob ~ is tailored to the fusion rules
[\bs][\bss].  The basic suggestion came to us from an interesting paper of
Cardy \Ref\cardy{J.L. Cardy, {\sl Nucl. Phys.} {\bf B324} (1989) 581.} on the
annulus amplitude in rational conformal field theory.  There, starting from
the correspondence between types of boundaries and bulk sectors present in
this case, the author relates the matter content flowing through the annulus
with boundaries of types $i$ and $j$ to the fusion-rule coefficient
$N_{ij}^k$.  Amusingly, if one extends Cardy's expression writing
$$
A \ =  \ {1 \over 2} \ \sum_{ijk} \ N_{ij}^k  \ n^i \ n^j \ \chi_k	\qquad ,
\eqn\acardy
$$
the Verlinde formula\Ref\ver{E. Verlinde, {\sl Nucl. Phys.} {\bf B300}
(1988) 360.}
implies that the coefficients of the
transverse-channel amplitude are perfect squares.  Indeed
$$
\tilde{A} \ \sim  \ \sum_{jm} \ {\bigg( {{S_j^m  n^j} \over {\sqrt{S_0^m}}}
\bigg)}^2 \chi_m	\qquad ,
\eqn\ver
$$
where $S$ denotes the matrix that implements on the characters of the conformal
theory the transformation $\tau \rightarrow - 1/ \tau$.  In
lower-dimensional models, this choice does not exhaust all
possibilities, and indeed different ones are available, even in rational
models, at the minor price of altering the definition of the M\"obius
characters [\bs][\bss].  The basic idea is that one has the freedom to redefine
the eigenvalues under ``twist'' of various Virasoro primaries within a
generalized character, while preserving well-defined local currents on the
double cover of the M\"obius strip. In
ref. [\bss], the resulting modifications were ascribed to
``discrete'' Wilson lines running along the boundaries of the annulus.
Amusingly, in toroidal models the different choices of ``discrete'' Wilson
lines are particular points of continuous lines of deformations associated with
internal gauge fields, and the name is thus properly justified.

One may wonder whether the construction based on eq. \acardy ~ is limited to
the case of simple models built out of free fermions and bosons on the world
sheet or, rather, whether it may be extended to models with a more
complicated fusion algebra.  As a first step in this direction, in ref.
\Ref\discrete{M. Bianchi, G. Pradisi and A. Sagnotti, {\sl Phys. Lett.}
{\bf 273B} (1991) 389.} we have shown that, starting from the discrete
series of Belavin, Polyakov and Zamolodchikov, one may derive
``open-string descendants'' whose disk
amplitudes enjoy proper factorization in the presence of a (global) internal
Chan-Paton symmetry.  In other words, one may construct Veneziano-type
analogues of the usual (Shapiro-Virasoro-like) amplitudes introduced in the
third of refs. [\clos].  It would be nice to extend this work to the other
discrete series and to $WZW$ models.

An important
issue is whether the massless spectrum has
some peculiar features that suffice to distinguish open-superstring models
from heterotic models.   In ten dimensions the answer is no, but this should
be regarded as an accident.  Indeed, in this case supersymmetry determines the
low-energy theory completely up to two derivatives, and identifies it in both
cases with $N=1$ supergravity coupled to $N=1$ supersymmetric Yang-Mills. The
next interesting case is $D=6$, and indeed in ref. [\bss] we
constructed a class of supersymmetric open-string models in six dimensions that
stands out by its peculiarities.  Following the suggestion of ref. [\car],
these models were derived starting from $K_3$ compactifications of the
type-IIb superstring and performing the parameter-space orbifold
construction.  They show the novel feature (compared to the heterotic
case) of containing a net number of self-dual tensors $B_{\mu \nu}$ in
their spectra.  Technically, these fields originate from incomplete
projections of the $R - R$ sector of the ``parent'' closed string, and thus
are certainly not present in the heterotic string.  Still, they play a very
important role, since they contribute to cancelling the irreducible part of
the anomaly polynomial.

Referring to fig. 8, let us first confine our attention to the gauge
anomaly in ten dimensions.  The simplest relevant diagrams contain six
external vectors and are of three types.  In the planar diagrams all six
vectors are emitted from one boundary of the annulus.  In this case there is a
potential singularity in the limit of very long tubes that is not regulated by
any momentum flow. A similar pathology is also present in the non-orientable
amplitudes, where all six vectors are emitted from the single boundary of the
M\"obius strip.  These two kinds of amplitudes determine the total irreducible
contribution to the anomaly polynomial, proportional in this case to
${\rm Tr} F^6$.  The tadpole condition removes this
contribution if the gauge group is $SO(32)$.  On the other hand,
the non-planar diagram is not singular by itself.  It involves the emission
of four vectors from one boundary and two vectors from the other boundary
of the annulus, and therefore is regulated by the momentum flow along the
tube. Still, from the viewpoint of the low-energy theory, this diagram is the
most important one, since it is the site of the Green-Schwarz mechanism,
whereby the $R-R$ antisymmetric tensor disposes of the residual anomaly.
We should add that the vacuum diagrams contain detailed information on
the cancellation mechanism for the irreducible part of the
gravitational
anomaly, proportional to ${\rm Tr} R^6$. In this case the relevant
contribution comes from the limiting behavior of the amplitude when all six
emission vertices coalesce, and thus one may relate the
phenomenon, once more, to the tadpole conditions.  Since, by
construction, they result in second-degree
polynomials with coincident roots, one finds again that $SO(32)$ does the job.
Indeed, the limiting behavior of the six-vector amplitudes is proportional to
$(N-32)$, whereas the total contribution of the vacuum amplitudes is
proportional to $(N-32)^2$. All
this should be compared to the corresponding mechanism in closed-string
theories, where modular invariance removes the ultraviolet singularity
altogether,
and is thus responsible for the whole anomaly cancellation.

In six dimensions, the anomaly polynomial contains irreducible terms
proportional to $\rm{Tr} F^4$ and $\rm{Tr} R^4$ that are disposed of by
the tadpole conditions.  They
are associated, respectively, with gauge and gravitational anomalies.  The
six-dimensional models of ref. [\bss] contain a number of different sectors,
and in particular a number of tadpole sectors.  Since each of the tadpole
sectors contains an antisymmetric tensor\foot{One of these is antiself-dual and
belongs to the supergravity multiplet, while the remaining ones
are self-dual and belong to the matter multiplets.}, one may wonder
whether they might all play a role in the cancellation mechanism.
Indeed, the analysis  carried out in
ref. [\as] revealed a novel feature: in this case the
Green-Schwarz mechanism does take a generalized form!  From a technical
viewpoint, one finds that, after imposing the tadpole conditions, the residual
anomaly polynomial is a quadratic form built out of $\rm{Tr} F^2$ and
$\rm{Tr} R^2$ that, in sharp contrast with the usual case, {\it does not}
factorize. Rather, the quadratic form may be diagonalized and contains a
number of nonzero eigenvalues.  Interestingly, this number is precisely equal
to the number of antisymmetric tensors in the models, that may thus dispose of
the anomaly by acting in a combined fashion.  The low-energy effective
supergravity should thus contain generalized couplings between
antisymmetric tensors and combinations of Chern-Simons forms for the various
simple factors of the gauge group, weighted by the matrix that implements on
the characters the transformation  $\tau \rightarrow - 1 /
\tau$.  The role of this matrix is precisely as expected, since in the
non-planar case the contributions to the anomaly polynomial are weighted by
the same factors as the contributions to the tadpole graphs.  This should
indeed be the case, since the same Feynman rules determine the GSO-type
contributions in both cases.  Thus, one may predict the canonical form of the
residual anomaly polynomial!  On the other hand, the gravitational
Chern-Simons form couples only to the antisymmetric tensor in the supergravity
multiplet. One may also write a supersymmetric set of field equations for
the low-energy field theory that incorporate these generalized Yang-Mills
couplings [\as].  Following common practice\Ref\sdual{ J.H. Schwarz, {\sl Nucl.
Phys.} {\bf B226} (1983) 289;\nextline L.J. Romans, {\sl Nucl. Phys.} {\bf
B276} (1986) 71.}, these equations have been constructed only to lowest order
in the spinors, but  previous experience  \Ref\hw{P.S. Howe and P.C. West,
{\sl Nucl. Phys.} {\bf B238} (1984) 181.} suggests that
their completion to all orders should entail only difficulties of a technical
nature.

In conclusion, we have seen how, in going from ten to six dimensions,
open-string models display an enticing complexity.  Of course, many major
questions remain unanswered.  First of all, we have
omitted any direct reference to chiral four-dimensional models. As far as we
know, open-string models appear to be plagued with a common disease: all chiral
four-dimensional models they give rise to involve small gauge groups,
typically products of $U(2)$ factors.  Whereas in some cases this finding may
be related to the need to introduce (quantized) background values of the
$NS-NS$ antisymmetric tensors [\torus], the argument is not exhaustive, and a
detailed analysis is called for.  An additional problem is the
geometrical formulation of these models.  For instance, a basic question has to
do with the residual antisymmetric tensors: how can one understand the
incomplete projections of $R-R$ sectors directly in terms of the type-I
superstring? Is the ``parameter-space orbifold'' more than an technical
artifice?  Hopefully, we shall be able to report on some of these
issues at a future Meeeting of this Society.

\vskip 50pt
\centerline{\bf Acknowledgments}

We are grateful to M. Bianchi, with whom several of the results reviewed in
this talk were derived, during a long and enjoyable collaboration.
A.S. would like to thank the Organizers for their kind
invitation, while apologizing for his suddend inability to
attend the Meeting.

\vskip 24pt

\vskip 24pt
\centerline{\bf Figure Captions}
\vskip 24pt
{\bf Figure 1}

The torus amplitude

{\bf Figure 2}

The annulus amplitude

{\bf Figure 3}

Genus-one vacuum amplitudes in open-string theories

{\bf Figure 4}

Vacuum channels

{\bf Figure 5}

Tadpole conditions

{\bf Figure 6}
Allowed sectors in the annulus vacuum channel

{\bf Figure 7}

Chan-Paton symmetry breaking

{\bf Figure 8}

The Green-Schwarz mechanism

\endpage
\refout
\end